# Permutation attack and counterattack on the two-party quantum key agreement over a collective noisy channel

Jun Gu[1], Tzonelih Hwang[*]

[1]*Department of Computer Science and Information Engineering, National Cheng Kung University, No. 1, University Rd., Tainan City, 70101, Taiwan, R.O.C.*

[1] isgujun@163.com

[*]**Corresponding Author:**

Tzonelih Hwang

Distinguished Professor

Department of Computer Science and Information Engineering,

National Cheng Kung University,

No. 1, University Rd.,

Tainan City, 70101, Taiwan, R.O.C.

Email: hwangtl@csie.ncku.edu.tw

TEL: +886-6-2757575 ext. 62524

# Abstract

Recently, Yang et al. (Quantum Inf Process 18, 74, 2019) proposed a two-party quantum key agreement protocol over a collective noisy channel. They claimed that their quantum key agreement protocol can ensure both of the participants have equal influence on the final shared key. However, this study shows that the participant who announces the permutation operation can manipulate the final shared key by himself/herself without being detected by the other. To avoid this loophole, an improvement is proposed here.

**Keywords** Quantum cryptography. Quantum key agreement. Permutation attack.

## 1. Introduction

Since the first quantum key distribution (QKD) protocol has been proposed in 1984 [1], several QKD protocols [2, 3] and related protocols [4, 5] have been proposed. However, in these QKD protocols, the shared secret key is first determined by a participant or a third-party and then transmitted to the other participants. Different from the QKD protocols, to ensure all the participants have equal influence on the shared key, the quantum key agreement [6] (QKA) was proposed. In a QKA, none of the proper subset of the involved participants can determine any part of the final shared key without being detected by the others.

Recently, Yang et al. [7] proposed a two-party QKA protocol over a collective noisy channel. They claimed that, in their QKA protocol, each participant has an equal contribution to the final shared key. But this study shows that Yang et al.'s QKA protocol suffers from a permutation attack. That is, the participant who announces the permutation operation can manipulate the final shared key without being detected. Then, a simple modification is hence proposed here.

The rest of this paper is organized as follows. In Section 2, Yang et al.'s QKA protocol is reviewed. In Section 3, we show the details of the permutation attack on Yang et al.'s QKA protocol and then propose a modified method to solve this problem.

At last, a conclusion is given in Section 4.

## 2. A brief review of Yang et al.'s QKA

Before reviewing Yang et al.'s QKA protocol [7], some background is introduced first here.

### 2.1 Background

In [8], four decoherence-free subspace (DFS) states $\{|0_{dp}\rangle=|01\rangle, |1_{dp}\rangle=|10\rangle,$ $|+_{dp}\rangle=\frac{1}{\sqrt{2}}(|0_{dp}\rangle+|1_{dp}\rangle), |-_{dp}\rangle=\frac{1}{\sqrt{2}}(|0_{dp}\rangle-|1_{dp}\rangle)\}$ against the collective-dephasing noise are presented. Similarly, to avoid the collective-rotation noise, four different DFS states $\{|0_r\rangle=\frac{1}{\sqrt{2}}(|00\rangle+|11\rangle), |1_r\rangle=\frac{1}{\sqrt{2}}(|01\rangle-|10\rangle), |+_r\rangle=\frac{1}{\sqrt{2}}(|0_r\rangle+|1_r\rangle), |-_r\rangle=\frac{1}{\sqrt{2}}(|0_r\rangle-|1_r\rangle)\}$ can be used.

In Yang et al.'s QKA protocol, the logical quantum states $\{|0_L\rangle, |1_L\rangle, |+_L\rangle, |-_L\rangle\}$ are used where $L$ represents '$dp$' or '$r$'. It means that if the states are used to avoid the collective-dephasing noise, $L$ represents '$dp$'. Similarly, the $L$ represents '$r$' if the states are transmitted via a quantum channel with the collective-rotation noise. According to the logical quantum states, four logical Bell states $\{|\Phi_L^+\rangle, |\Phi_L^-\rangle, |\Psi_L^+\rangle, |\Psi_L^-\rangle\}$ can be described as follows:

$$\begin{cases} |\Phi_L^+\rangle = \frac{1}{\sqrt{2}}(|0_L 0_L\rangle + |1_L 1_L\rangle) \\ |\Phi_L^-\rangle = \frac{1}{\sqrt{2}}(|0_L 0_L\rangle - |1_L 1_L\rangle) \\ |\Psi_L^+\rangle = \frac{1}{\sqrt{2}}(|0_L 1_L\rangle + |1_L 0_L\rangle) \\ |\Psi_L^-\rangle = \frac{1}{\sqrt{2}}(|0_L 1_L\rangle - |1_L 0_L\rangle) \end{cases} \quad (1)$$

Moreover, four logical unitary operations $\{U_{00}^L, U_{01}^L, U_{10}^L, U_{11}^L\}$ are used in Yang et al.'s QKA protocol. Each logical Bell state can be transformed into another logical Bell state by performing the logical unitary operations and the transformation relationships are shown in Table 1.

Table 1. The Bell states transformation results

| Initial state | $U_{00}^L$ | $U_{01}^L$ | $U_{10}^L$ | $U_{11}^L$ |
| --- | --- | --- | --- | --- |
| $\|\Phi_L^+\rangle$ | $\|\Phi_L^+\rangle$ | $\|\Phi_L^-\rangle$ | $\|\Psi_L^+\rangle$ | $\|\Psi_L^-\rangle$ |
| $\|\Phi_L^-\rangle$ | $\|\Phi_L^-\rangle$ | $\|\Phi_L^+\rangle$ | $\|\Psi_L^-\rangle$ | $\|\Psi_L^+\rangle$ |
| $\|\Psi_L^+\rangle$ | $\|\Psi_L^+\rangle$ | $\|\Psi_L^-\rangle$ | $\|\Phi_L^+\rangle$ | $\|\Phi_L^-\rangle$ |
| $\|\Psi_L^-\rangle$ | $\|\Psi_L^-\rangle$ | $\|\Psi_L^+\rangle$ | $\|\Phi_L^-\rangle$ | $\|\Phi_L^+\rangle$ |

The Bell state entanglement swapping [9] also is used in Yang et al.'s QKA protocol. That is, suppose that $IS_1, IS_2$ are the initial states of two logical Bell states, respectively. Here, $IS_i \in \{|\Phi_L^+\rangle = \text{'00'}, |\Phi_L^-\rangle = \text{'01'}, |\Psi_L^+\rangle = \text{'10'}, |\Psi_L^-\rangle = \text{'11'}\}$, $i \in \{1, 2\}$. By performing logical Bell measurement on both of the first logical particles and both of the second logical particles we can obtain the measurement results $MR_1$ and $MR_2$, respectively. The entanglement swapping of these logical Bell states follows the following formula

$$MR_1 \oplus MR_2 = IS_1 \oplus IS_2 \qquad (2)$$

**2.2 Yang et al.'s QKA protocol**

Suppose that there are two participants Alice and Bob involved in Yang et al.'s QKA protocol and they have $2n$ bits secret key $K_A = \{k_A^1, k_A^2, \cdots, k_A^n\}$ and $K_B = \{k_B^1, k_B^2, \cdots, k_B^n\}$, respectively. Here, $k_A^i, k_B^i \in \{00, 01, 10, 11\}$ $(1 \leq i \leq n)$. Then the protocol can be described step by step as follows.

**Step 1.** Alice generates $2n$ logical Bell states in $|\Phi_L^+\rangle$, and picks out all the first logical particles and all the second logical particles of $|\Phi_L^+\rangle$ to form the ordered logical particle sequence $S_A=\{s_A^1, s_A^2, \cdots, s_A^{2n}\}$ and $S_B=\{s_B^1, s_B^2, \cdots, s_B^{2n}\}$, respectively. After that, Alice generates enough logical decoy photons where each logical photon is randomly selected from $\{|0_L\rangle, |1_L\rangle, |+_L\rangle, |-_L\rangle\}$ and randomly inserts these decoy photons into $S_B$ to obtain a new sequence $S_B'$. Finally, Alice sends $S_B'$ to Bob.

**Step 2.** After confirming that Bob has received $S_B'$, Alice and Bob use the logical decoy photons to check whether there is an eavesdropper during the particle transmission process or not. That is, Alice first announces the positions and the initial states of the logical decoy photons in $S_B'$, then Bob uses the corresponding basis to measure each decoy photon. If the error rate of the measurement results exceeds a predetermined value, this protocol will be aborted. Otherwise, the protocol will go to the next step.

**Step 3.** After removing all the decoy photons, Bob can obtain the particle sequence $S_B$. Subsequently, Alice and Bob perform logical Bell measurement on the particles $\{s_A^{2i-1}, s_A^{2i}\}$ and $\{s_B^{2i-1}, s_B^{2i}\}$ $(1\leq i \leq n)$ to obtain the measurement result sequence $M_A=\{m_A^1, m_A^2, \cdots, m_A^n\}$ and $M_B=\{m_B^1, m_B^2, \cdots, m_B^n\}$, respectively. According to the formula (2), we can know that $M_A=M_B$. Here, for simplicity, $M=\{m^1, m^2, \cdots, m^n\}$ is used to represent both of $M_A$ and $M_B$.

**Step 4.** Alice generates $n$ logical Bell states according to $M$ and performs one of the logical unitary operations $\{U_{00}^L, U_{01}^L, U_{10}^L, U_{11}^L\}$ on each Bell state according to $K_A$. The generated logical Bell state sequence is named as $S_C$. Then Alice performs a random permutation operation $\Pi_n$ on $S_C$ to obtain $S'_C$. At last, Alice inserts the logical decoy photons into $S'_C$ the same as she did in Step 1 to obtain $S''_C$ and sends $S''_C$ to Bob.

**Step 5.** After Bob receives $S''_C$, Alice and Bob use the logical decoy photons in $S''_C$ to check the eavesdropping the same as they did in Step 2. Then Bob can obtain $S'_C$.

**Step 6.** Bob announces the value of $K_B \oplus M = \{k_B^1 \oplus m^1, k_B^2 \oplus m^2, \cdots, k_B^n \oplus m^n\}$. Hence, Alice can derive Bob's secret key $K_B = \{k_B^1, k_B^2, \cdots, k_B^n\}$ by the equation $K_B \oplus M \oplus M = K_B$.

**Step 7.** Alice publishes the permutation operation $\Pi_n$ so that Bob can perform an inverse permutation operation on $S'_C$ to get $S_C$. Then Bob measures $S_C$ with the logical Bell measurement to obtain $M \oplus K_A$ and derives Alice's secret key $K_A$.

**Step 8.** Alice and Bob compute the final shared key $K_{AB} = (K_A \oplus K_B) \| (K_A \oplus K_B \oplus M)$.

## 3. A loophole in Yang et al.'s QKA protocol and an improvement

Yang et al. claimed that the above QKA protocol can ensure both Alice and Bob have equal contribution to the final shared key $K_{AB}$ and none of them can manipulate $K_{AB}$ without being detected by the other. However, this section shows that Alice can announce a fake permutation operation to select a preferred final key $K'_{AB}$ instead.

Then, to solve this problem, a simple solution is proposed.

## 3.1 The loophole in Yang et al.'s QKA protocol

At the end of Step 6, Alice can obtain Bob's secret key $K_B$ and then computes the final shared secret key $K_{AB}=(K_A \oplus K_B) \| (K_A \oplus K_B \oplus M)$. If she does not want to use $K_{AB}$ as the final shared secret key, then she can announce a fake permutation operation $\Pi'_n$ in Step 7 instead. Upon receiving the fake permutation operation, Bob will use a corresponding fake inverse permutation operation to reorder $S'_C$ to a fake final key $K'_{AB}=(K'_A \oplus K_B) \| (K'_A \oplus K_B \oplus M)$ which is chosen by Alice.

For example, assume that $K_A=\{k_A^1, k_A^2\}=0011$, $K_B=\{k_B^1, k_B^2\}=0110$ and $M=\{m^1, m^2\}=1110$, then the corresponding final key will be $K_{AB}=(K_A \oplus K_B) \| (K_A \oplus K_B \oplus M)=01011011$. With the permutation attack, if Alice wants to choose a fake final key 11110001 instead, she can announce a fake permutation operation which reorders the measurement results of $S_C$ from $K_A \oplus M=\{k_A^1 \oplus m^1, k_A^2 \oplus m^2\}=1101$ to $(K_A \oplus M)'=\{k_A^2 \oplus m^2, k_A^1 \oplus m^1\}=0111$. Then, Bob will obtain a fake $K'_A=(K_A \oplus M)' \oplus M=1001$ and gets the fake final key $K'_{AB}=(K'_A \oplus K_B) \| (K'_A \oplus K_B \oplus M)=11110001$.

## 3.2 A solution to the loophole

Because all the photons of $S_C$ have been reordered, Alice can reorder these logical Bell states into a preferred order to manipulate the final shared key. If the permutation operation just performed on all the first logical qubits of the logical Bell states in $S_C$, without the correct permutation operation, the Bell measurement performed by Bob on the particles will result in an entanglement swapping. The entanglement swapping makes the measurement results random and hence Alice has no motivation to announce the fake permutation operation other than disrupting the QKA protocol. The detail of the improvement is as follows.

**Step 1\*-3\*** are the same as **Step 1-3** in Section 2.

**Step 4\*.** Similarly, Alice generates the logical Bell state sequence $S_C$. Then Alice performs a random permutation operation $\Pi_n$ on all the first logical particles of Bell state in $S_C$ to obtain $S'_C$. At last, Alice inserts the logical decoy photons into $S'_C$ as same as she does in Step 1 to obtain $S''_C$ and sends $S''_C$ to Bob.

**Step 5\*-8\*** are the same as **Step 5-8** in Section 2.

With this simple modified method, the permutation attack can be avoided.

## 4. Conclusions

Yang et al. proposed a two-party quantum key agreement protocol over a collective noisy channel. However, this study shows that Yang et al.'s QKA protocol suffers from a permutation attack. A solution is hence proposed here to avoid the loophole.

## Acknowledgment

We would like to thank the Ministry of Science and Technology of the Republic of China, Taiwan for partially supporting this research in finance under the Contract No. MOST 107-2627-E-006-001; No. MOST 108-2221-E-006-107.